\documentclass[10pt, conference]{IEEEtran}

\makeatletter
\def\footnoterule{\relax%
  \kern-5pt
  \hbox to \columnwidth{\hfill\vrule width 0.8\columnwidth height 0.4pt\hfill}
  \kern4.6pt}
\makeatother

\usepackage{graphicx}
\usepackage[utf8]{inputenc}

\usepackage{multirow}
\usepackage[utf8]{inputenc}
\usepackage{lipsum}
\usepackage{multicol}
\usepackage{graphicx}

\usepackage{comment} 

\usepackage{float}
\usepackage{dblfloatfix}

\usepackage{listings}
\usepackage{xcolor}
\usepackage{hyperref}
\let\svthefootnote\thefootnote
\newcommand\freefootnote[1]{%
  \let\thefootnote\relax%
  \footnotetext{#1}%
  \let\thefootnote\svthefootnote%
}

\newcommand{\FIXME}[1]{{\color{red} \bf  #1}}

\title{Designing a Streaming Data Coalescing Architecture for Scientific Detector ASICs with Variable Data Velocity}
\author{
\IEEEauthorblockN{Sebastian Strempfer, Kazutomo Yoshii, Mike Hammer, Dawid Bycul, Antonino Miceli}
\IEEEauthorblockA{Argonne National Laboratory\\Lemont, IL}
}

\begin{document}
\maketitle

\section{Abstract}
Scientific detectors are a key technological enabler for many disciplines. Application-specific integrated circuits (ASICs) are used for many of these scientific detectors. Until recently, pixel detector ASICs have been used mainly for analog signal processing of the charge from the sensor layer and the transmission of raw pixel data off the detector ASIC. However, with the availability of more advanced ASIC technology nodes for scientific application, more digital functionality from the computing domains (e.g., compression) can be  integrated directly into the detector ASIC to increase data velocity. However, these computing functionalities often have high and variable latency, whereas scientific detectors must operate in real-time (i.e., stall-free) to support continuous streaming of sampled data. This paper presents an example from the domain of pixel detectors with on-chip data compression for X-ray science applications. To address the challenges of variable-sized data from a parallel stream of compressors, we present an ASIC design architecture to coalesce variable-length data for transmission over a fixed bit-width network interface.

\begin{IEEEkeywords}
Scientific instrument edge systems, X-ray science, data transfer technologies, streaming data compression, X-ray detectors, ASIC, hardware construction languages

\end{IEEEkeywords}


\section{Introduction}

\begin{figure*}[!hb]
  \centering 
  \includegraphics[width=.9\textwidth,origin=c,angle=0]{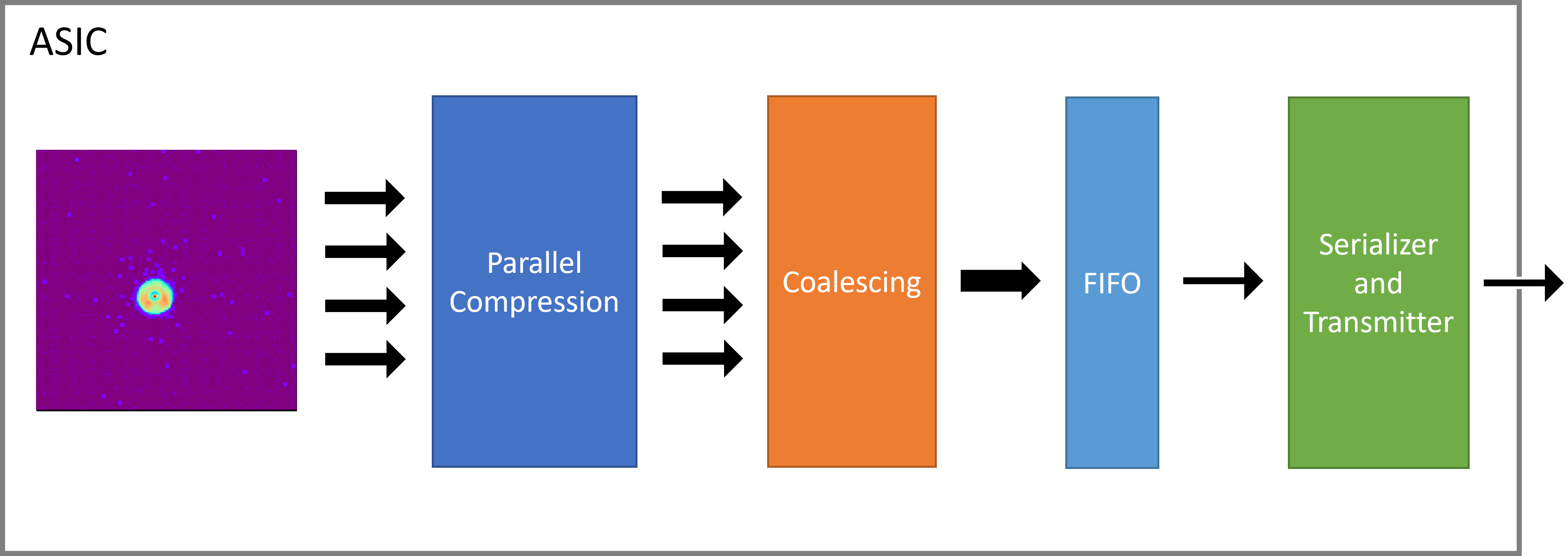}
  \caption{Data Acquisition System (DAQ) on our detector chip. The image shown is a false-color X-ray ptychography image.}
  \label{fig:detector-overview} 
\end{figure*}

Pixel detectors are at the heart of advances in imaging. Astronomy has been revolutionized by the use of high-throughput detectors for surveying supernovas, shedding light on expansion rates of the universe~\cite{Riess_1998}, and exoplanet searches using the wobble of faint stars~\cite{Mayor:1995cp}. Biochemistry has been transformed first by developing megapixel detectors for X-ray crystallography~\cite{CCD}, and more recently, by using high frame rate detectors to correct for drift and enable atomic resolution structure determination in cryo-electron microscopy~\cite{Li:2013gt}. Materials scientists are gaining new insights into subtle electron bonding arrangements using high dynamic range electron detectors as part of phasing coherent diffraction patterns~\cite{Jiang:2018jna}.  Beyond scientific advances, the photography industry has been revolutionized by the transition to electronic image detectors~\cite{Fossum}. 

The scientific pixel detectors are typically built using application-specific integrated circuits (ASICs). In general, integrated circuits are at the core of smartphones and cameras, the internet of things, computers, cloud data centers, and a new wave of artificial intelligence processors. What has enabled the proliferation of integrated circuits beyond traditional computing units (i.e., CPUs) is the “pure-play” semiconductor foundry business model, where even small-scale companies can develop designs that are then fabricated using billion-dollar semiconductor fabrication facilities. This resulted in a new opportunity for fabless semiconductor companies to emerge which focus only on integrated circuit design. The academic research community is fortunate to have access to these foundries through third-party brokers who facilitate cost-effective prototyping of small designs and full-scale fabrications at  advanced technology nodes. With the availability of these advanced technology nodes, the number of digital logic resources available increases. The use of digital circuitry also opens up the possibility of incorporating  digital data processing (e.g., compression, machine learning, etc) directly on the detector ASIC.

Future scientific breakthroughs depend on more comprehensive observations, translating into increased spatial and temporal resolution in scientific measurements. Ever-increasing data velocity from scientific instruments is becoming a considerable challenge in the network connection between detector chips and a data acquisition system. Unlike server systems, upgrading the network connection between them requires further complex I/O designs on both ends, not simply replacing network adapters and cables. Another approach is to design an on-chip data compressor or data analysis logic for detector chips to reduce the data size. While the data generated by sensors is fixed in size (e.g., the camera resolution is fixed), the compressed data block is variable. The challenge then is that the network interface bit-width is fixed. For example, the detector chip design presented in this paper sends a fixed 512-bit word to the network interface.

It is critical to pack multiple variable-sized compressed blocks together to fill the network interface efficiently; otherwise, it lowers the network utilization. However, designing such a data packing mechanism is challenging because of higher resource usages. For example, the variable to fixed length converter in a bandwidth compressor design consumes a large percentage (e.g., \textgreater80\%) of the entire resources~\cite{Ueno:2017hv}. It is also challenging to optimize the performance of such implementations, particularly for high-speed designs and designs for field-programmable gate arrays (FPGAs). Although data packing mechanisms are mandatory, they tend to be underrated. More effort is typically focused on developing novel compression algorithms, leading to implementation of one-off packing designs. This paper describes the details of our data coalescing architecture with the intent of laying the groundwork for a reusable design.

\section{Background}

Previously, we have investigated the characteristic of X-ray data using four selected experimental X-ray data sets (XRD, ptychography, XPCS concentrated and dilute) in order to design hardware algorithms that can reduce the amount of data from pixel arrays~\cite{Hammer_2021}. In our previous study, we primarily focused to study compressor hardware algorithms that leverage statistical redundancy in data, which can be placed in parallel to cover the entire image (the 'Parallel Compression' stage in Figure~\ref{fig:detector-overview}). We have successfully implemented our hardware compressor algorithms in the Chisel hardware construction language~\cite{Bachrach:ft} and verified the functionality of the generated RTL code with both synthetic and actual X-ray datasets using Verilator~\cite{snyder2013verilator} via Chisel's test harness. 
However, in our previous study, no hardware logic that combines variable-sized compressed data from parallel compression blocks was discussed. This paper follows up our previous study covering on-chip compression and discusses our idea of a data stream coalescing architecture (the 'Coalescing' stage in Figure~\ref{fig:detector-overview}).

In the post-Moore era, hardware specialization plays an important role not only to improve performance but also to meet special requirements such as single-cycle, stall-free operation, which may be impossible with a system with general-purpose processors due to complex memory and I/O hierarchy, which are typically optimized for throughput. Our proposed solution is realized by means of hardware specialization and can also be an essential component in other specialized hardware designs if we provide our design as a reusable hardware library. In our recent study~\cite{YoshiiCCSI}, we discussed the importance of streaming computing at the edge of scientific instruments along with the challenges and opportunities within hardware specialization methodology. This included several usage examples of open-source hardware design tools such as Chisel and Yosys~\cite{wolf2013yosys}. Our proposed data coalescing architecture and its testbench are also written in Chisel and are available in GitHub as open-source software\footnote{\url{https://github.com/SEBv15/compression-reduction/tree/paper}}. Since Chisel is a domain-specific language that is implemented in Scala, a modern powerful general-purpose language, and fully integrates an RTL-level simulator environment, it is an adequate language to express hardware libraries in a more software-friendly manner. In fact, many real-world hardware libraries and architectures are written in Chisel~\cite{Asanovic:EECS-2016-17,celio2017boomv2,alon2018craft, lee2016agile}.

Pixel detectors are a critical component of physics experiment systems and have been studied for multiple decades~\cite{KeckPAD, ePix10k, AGIPD, LPD}. Their designs are shifting towards digital from analog in order to improve transmission rates and decrease transmission errors. Even though the data path between a detector chip and a data acquisition system is becoming a bottleneck as spatial and temporal resolution increases. Few works of literature discuss on-chip streaming data compressors and data coalescing designs, in particular stall-free designs, as of this writing. For example, a low-power high-speed readout design that leverages a binary tree priority encoder can achieve a frame rate of 1,000 Hz on a 64x64 pixel array~\cite{fahim2019low}.

\section{Design}

\begin{figure}[ht]
  \centering 
  \includegraphics[width=\columnwidth,origin=c,angle=0]{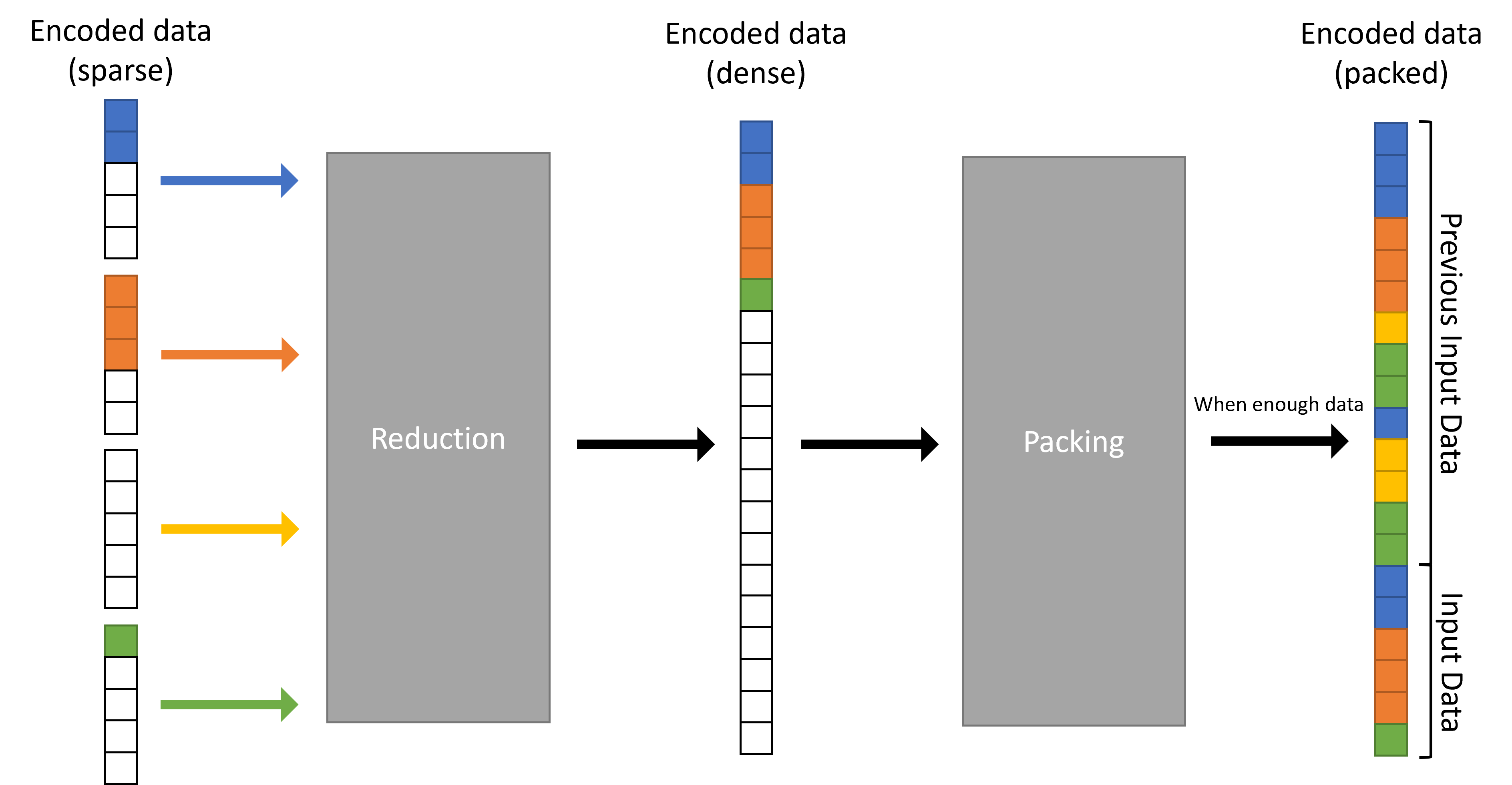}
  \caption{Conceptual layout of the coalescing module}
  \label{fig:coalescing-overview} 
\end{figure}

\begin{figure}[ht]
  \centering 
  \includegraphics[width=\columnwidth,origin=c,angle=0]{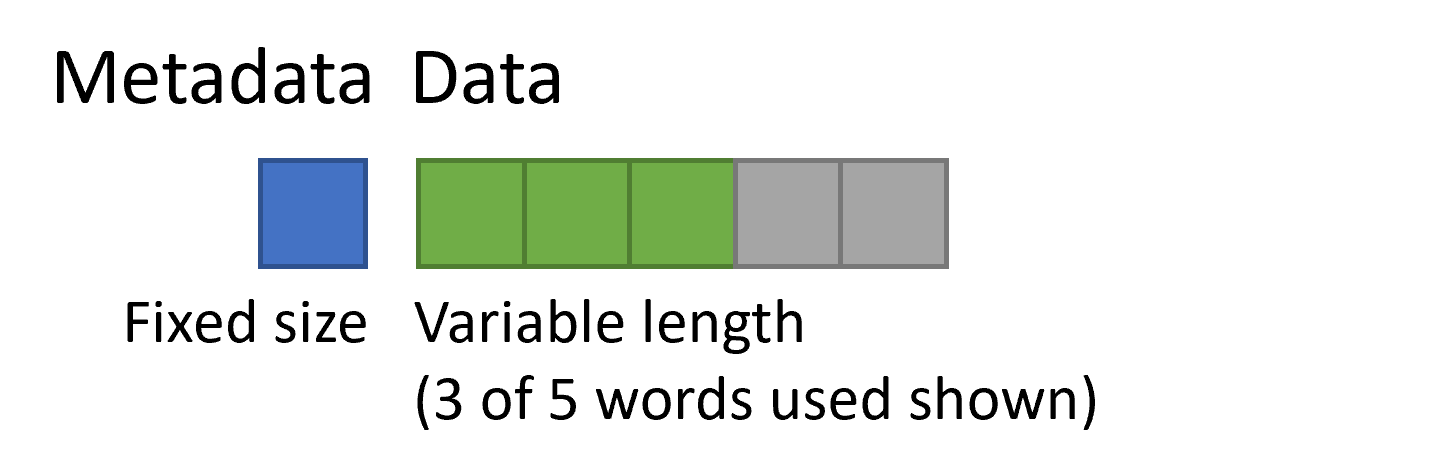}
  \caption{The output of a single encoder. Boxes represent a fixed size word.}
  \label{fig:encoder-output} 
\end{figure}

\begin{figure*}[htp]
  \centering 
  \includegraphics[width=.9\textwidth,origin=c,angle=0]{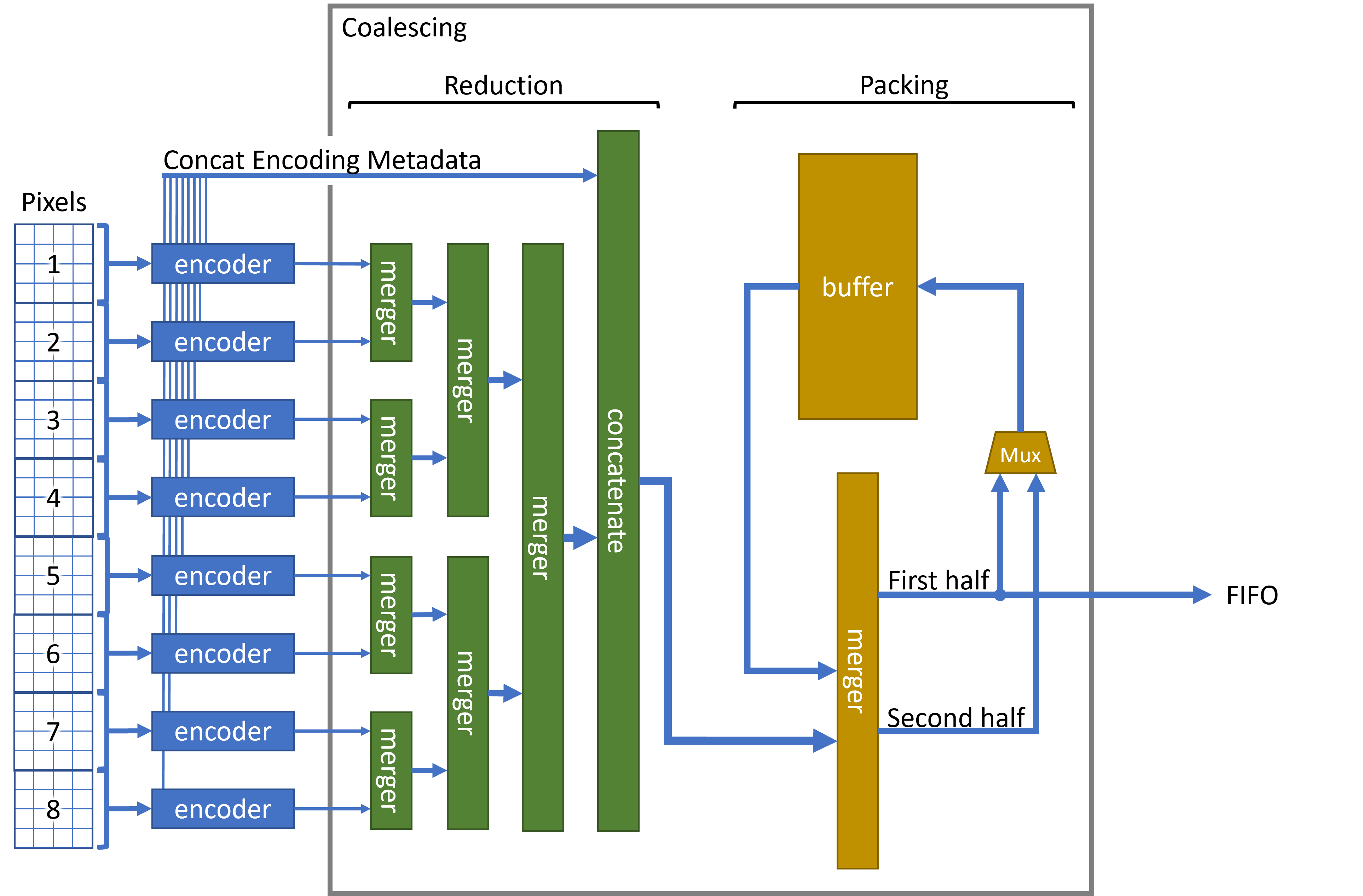}
  \caption{Conceptual layout of X-ray detector compression with eight parallel encoders}
  \label{fig:compressor-overview} 
\end{figure*}

\begin{figure*}[htbp]
  \centering 
  \includegraphics[width=.7\textwidth,origin=c,angle=0]{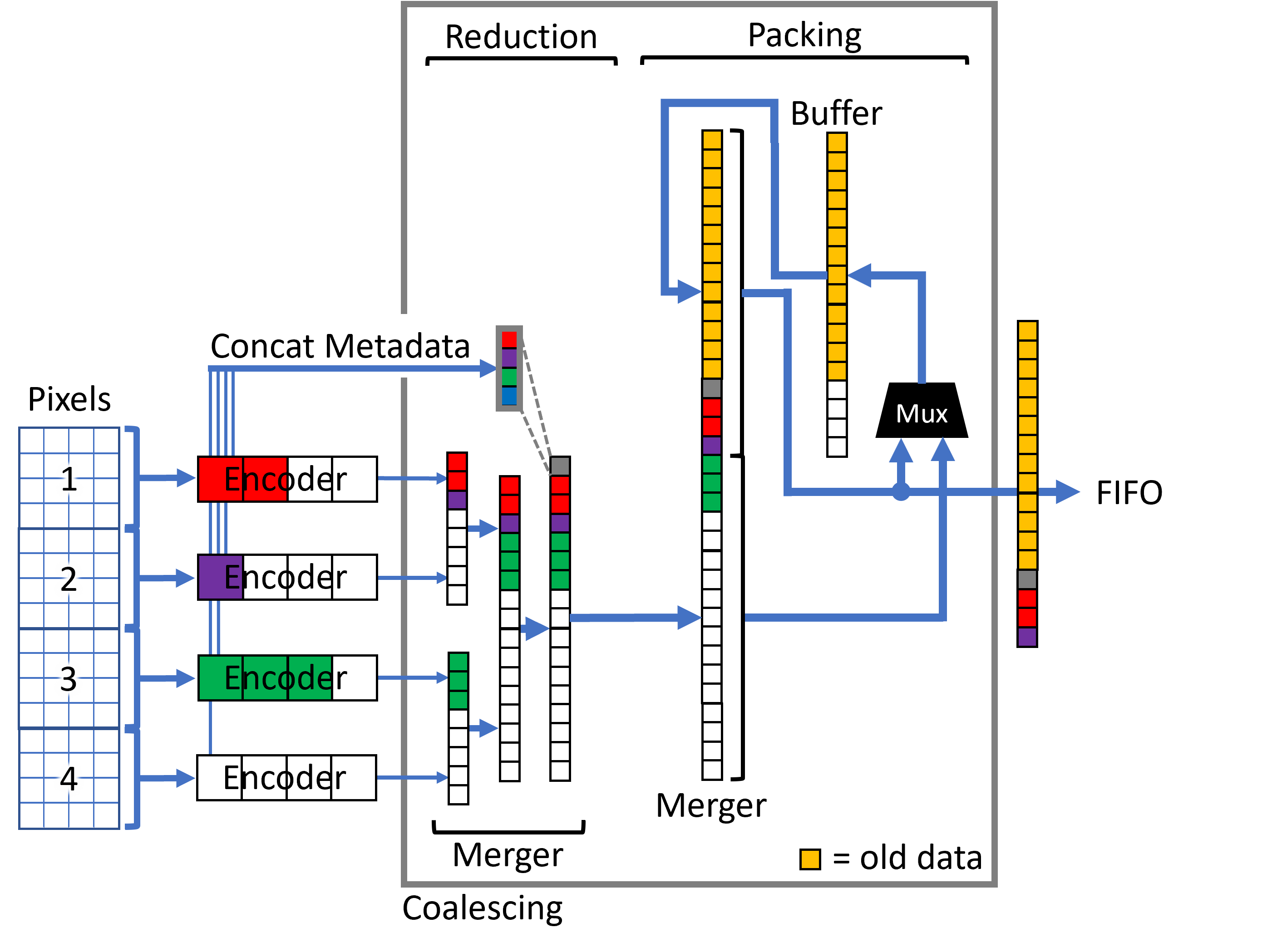}

  \caption{Conceptual layout of X-ray detector compression using four parallel encoders with example data. In this example, the green data will be stored in the \textit{Buffer} during the next clock cycle.}
  \label{fig:compressor-overview-data} 
\end{figure*}

The coalescing architecture consists of two distinct stages: reduction and packing (Figure \ref{fig:coalescing-overview}). The reduction stage takes the output of multiple parallel encoders, consisting of both constant and variable-length data (Figure \ref{fig:encoder-output}), and turns it into a single continuous variable-length output. The packing module then combines this variable-length data across clock cycles until a certain number of bits has accumulated. When ready, this constant-sized data is then passed on to the FIFO buffer and then the transmitter.

The packing is needed to ensure every packet of data the transmitter sends contains only useful bits. If the reduction output were used directly, transmitter packets might end up only partially filled. Even with packing, the entire bandwidth of the transmitter will most likely not be used since it is highly unlikely that the average data volume matches the transmission rate. However, using the packing module together with the FIFO, the transmitter bandwidth puts a lower limit on the average encoding efficiency. Without it, the limit would be applied to instantaneous efficiency. Using the FIFO and packing, variations in the reduction output size that would otherwise lead to data loss are now buffered and averaged out.

The conceptual layout of the coalescing logic used in an X-ray detector is shown in Figure \ref{fig:compressor-overview}. 
In this example, the data is being produced by a pixel array, shown on the left, and encoded using eight independent encoding modules. For this application, the packing stage and FIFO are especially necessary since there might be short events with a very bad compression ratio which need to be transmitted.

For off-chip transmission, it is necessary to pack this input into fixed-size words of data which can be sent one at a time.
In Figure \ref{fig:compressor-overview} it is shown how the encoder output is first merged into a single variable-length set of data in the reduction stage. This output (shown in Figure \ref{fig:encoder-output}) is then passed on to the packing module, which packs it into blocks across clock cycles.

Besides the buffer in the packing module, the coalescing logic is entirely combinational. This choice was made to simplify the design since it allows the compressor to use the same clock frequency as the rest of the system. In general, having a single clock domain is simpler for device implementation.
Using mostly combinational logic and a low clock frequency comes with the drawback of a larger physical chip area. However, this was only a secondary concern in the design of this architecture. Without the need for parallel encoders, the reduction module, which constitutes the majority of the logic, would be completely eliminated.

\subsection{Reduction}

\begin{figure}[!h]
  \centering 
  \includegraphics[width=\columnwidth,origin=c,angle=0]{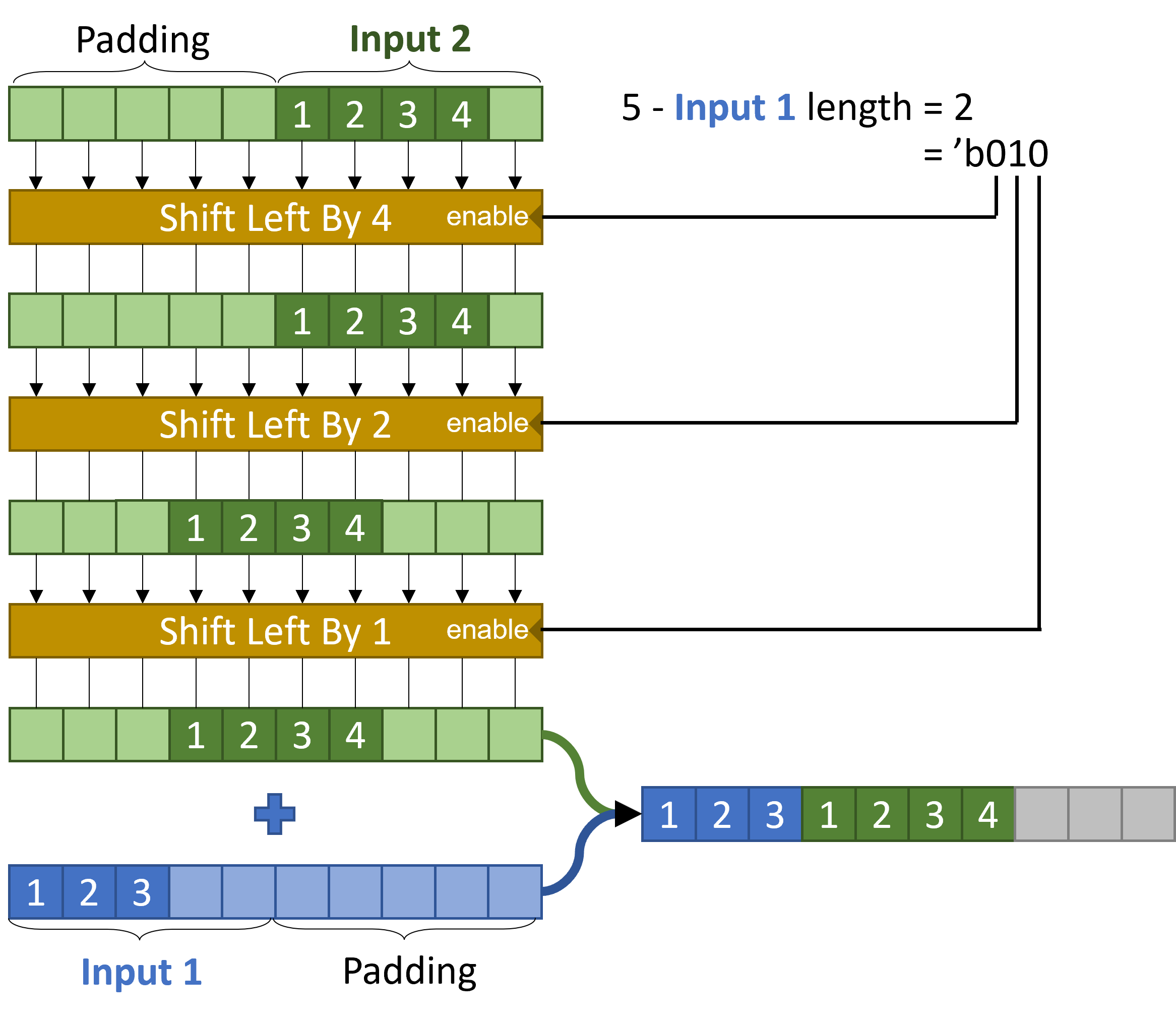}
  \caption{Merge module for two variable-sized inputs with a maximum size of 5 words}
  \label{fig:merger} 
\end{figure}

The transformation from multiple variable-sized data sets to a single output is referred to as reduction. It
is done by merging the variable-sized data in $\log_2{N}$ stages, 
where $N$ is the number of data sources.

Each stage combines two adjacent sets of variable-length data into a 
single variable-length output (shown in green in Figure \ref{fig:compressor-overview}).
A gate-efficient implementation of these "merge modules" is achieved by shifting
the second input by powers of two, in stages, until the two inputs occupy adjacent positions in the merged output (Figure \ref{fig:merger}).
The physical size of the module scales with $N\log{N}$ where $N$ is the maximum number of words in the variable-length data ($N=5$ in Figure \ref{fig:merger}).

To combine $M$ variable-length inputs into a single output then requires $\log_2(M)$ 
stages of merge modules. Each stage merges two adjacent inputs in a binary tree layout.
Therefore, the size of the entire reduction module scales with $MN\log(M)\log(N)$, 
where $M$ is the number of data sources and $N$ is the maximum number of words per source.

The constant-sized metadata output of the encoders can simply be concatenated
using wires and prepended to the output of the last reduction stage. Handling that output
separately from the variable-sized output is necessary to decrease the size of the merge modules. In addition, the merge modules rely on the data being
a multiple of a fixed wordsize long, meaning they cannot handle an arbitrary number of bits. Since the wordsize of the metadata is most likely different from that of the variable-length data, it must be processed separately. As pictured in Figure \ref{fig:compressor-overview-data}, the output of the reduction then consists of the constant-sized metadata (grey box) followed by the merged variable-length data.

\subsection{Packing}

After the reduction logic, the data is all in a single variable-length output. However, most transmitters work at a fixed rate and are only able to send a certain number of bits for every clock cycle. We are assuming that the encoders perform well enough so that their output bit-rate is less than that of the transmitter on average, but it is unrealistic to expect the same output size every time. This means there might be cycles where we have more data than we can send. To accommodate this influx, it is necessary to combine the reduced data across clock cycles and utilize a FIFO to even out the peaks in the data length.

This packing is done by utilizing a buffer. The output of the reduction module is merged with the buffer contents and stored back into it until enough data has accumulated to write the buffer contents to the FIFO and start over. The buffer size is determined by the maximum size of the reduction output since the packing logic needs to be able to handle the worst-case encoder performance. Since multiple reduction outputs might not fit evenly into the buffer, any leftover data can be written to the start of the buffer during the FIFO write. Figure \ref{fig:compressor-overview-data} shows what happens when there is enough data after merging the reduction output with previous data for a FIFO write.

Since this is a streaming output of variable-length encoded data, an error in transmission
would most likely be not recoverable. To avoid this, metadata can be included
at the beginning of every block to help find the start of a reduction output.

\section{Implementation and Verification}


We have implemented and verified our streaming data coalescing architecture using Chisel, a hardware construction language~\cite{Bachrach:ft,heilmann2013investigate,schoeberl2020digital}. Chisel is a domain-specific language for designing and simulating digital circuits. It is provided as a class library written in Scala, a general-purpose programming language that supports both object-oriented and functional programming~\cite{odersky2008programming}. Chisel generates synthesizable Verilog code~\cite{golson2016language} from digital circuit designs described in Chisel. It provides built-in test harnesses such as a peek-poke tester, which simulates input signals to the circuit and then compares the outputs to their expected values. Circuit designers can write test benches in Scala to verify the functionality of their designs at an RTL level, which allows us to create complicated test benches in a concise manner, thanks to the power of modern general-purpose programming languages. Chisel has a built-in RTL simulator and can interface with external simulators such as Verilator \cite{snyder2013verilator}, an open-source Verilog simulator.


We estimate the gate-level resource usage of the reduction and packing logic using iVerilog~\cite{iverilog} and Yosys~\cite{wolf2013yosys}. It is impossible to obtain a single absolute number on the die area of a circuit since it varies depending on many different factors such as optimization in synthesis tools, transistor feature size, and cell library. However, these tools can be utilized to obtain the circuit's relative sizes. iVerilog can quickly provide estimates by showing the gate count of a hardware circuit as-is. Conversely, Yosys is an open-source synthesis tool that optimizes the circuit before giving a gate count or area, if provided with logic gate sizes. In our analysis, both were used to compare iterations of our circuit and find the comparatively smallest viable version. 

Since the reduction module makes up the majority of the physical area of the coalescing logic, it was especially important to find an implementation that is as small as possible and scales reasonably well. Using the methods we just described, we confirmed that the reduction module area is proportional to $M\log{M}$, where $M$ is the number of parallel encoders.

\section{Conclusion}

We have demonstrated a design that is able to pack multiple variable-length inputs into constant-sized packets of data. The mostly combinational nature of the logic allows for a high throughput at low clock frequencies by taking in data every clock cycle. In the context of a high frame rate X-ray detector, this enables on-chip parallel data compression at the same clock speed as the pixel array readout. This gives us confidence that on-chip compressors are feasible in a streaming detector.

In the future, our efforts will be focused on two main goals. First, improving the existing codebase to make it more parameterizable, simplifying the reuse in future projects. Second, investigate how running the coalescing logic at a slightly higher clock frequency than the input and output would affect circuit area, power usage, and ease of implementation.

\section*{Acknowledgments}

This work is based on work supported by the U.S. Department of Energy, Office of Science, under contract  DE-AC02-06CH11357. This research is supported by Laboratory Directed Research and Development (LDRD 2021-0072) funding from Argonne National Laboratory.

\bibliographystyle{ieeetr}
\bibliography{references}

\clearpage

\section*{Appendix: Reproducibility}

The logic described in this paper is implemented in the compressor for our X-ray detector design. The source code for which can be found at \url{https://github.com/SEBv15/compression-reduction/tree/paper}. In addition to the coalescing logic, it also contains modules that are only relevant to the compression itself.

The logic is written entirely in Chisel, from which Verilog can be generated for further use.

\subsection{Requirements}

Since Chisel is a Scala library, the Scala build tool sbt\footnote{\url{https://www.scala-sbt.org/1.x/docs/Setup.html}} is needed to generate Verilog files and run the testbench. This is the only requirement to generate Verilog from the source code, however the testbench needs Verilator\footnote{\url{https://verilator.org/guide/latest/install.html}} to simulate the design.

To estimate the size of the modules from the generated Verilog, a Verilog synthesis tool is needed. A simple and lightweight choice is iVerilog\footnote{\url{https://iverilog.fandom.com/wiki/Installation_Guide\#Installation_From_Premade_Packages}}.

\subsection{Coalescing}

The coalescing module as described in this paper can be found under the source files as \texttt{Coalescing.scala}. The class arguments can be used to adjust the design parameters including number of parallel inputs, variable-length data size, etc.

The Verilog representation of the coalescing module can be generated using \texttt{sbt \textquotesingle runMain compression.Coalescing\textquotesingle}. 

The repository also contains a testbench for the coalescing module which can be found in the test directory as \texttt{CoalescingTest.scala}. It inserts random data into the coalescing module and checks if the output matches what is shown in Figure \ref{fig:compressor-overview-data}. It can be run using \texttt{sbt \textquotesingle testOnly compression.CoalescingTest\textquotesingle}.

\subsection{Size and scaling estimates}

Once the Verilog is generated from the Chisel source code, iVerilog can be used to compare the gate count of the module with different parameters\footnote{\texttt{iverilog -tsizer -o ivstat.txt Coalescing.v \&\& grep \textquotesingle Logic\textquotesingle \ ivstat.txt | tail -1}}. By generating the Coalescing module with different numbers of inputs and comparing their gate counts, the scaling of the coalescing logic can be demonstrated. The same can be done for the \texttt{Merge} and \texttt{Reduction} modules which should confirm the $N\log{N}$ and $NM\log{N}\log{M}$ scaling described in the paper.




\end{document}